\documentstyle[12pt]{article}
\begin{document}  
\vskip 2mm
\begin{center}
{\large\bf Renorm-group, Causality and Non-power \\ Perturbation Expansion
 in QFT \footnote{ To appear in the April issue of {\it Teor. Mat. Fizika}
  v.{\bf 119} (1999) No.1}
}
\vspace{4mm}

{\large D.V. Shirkov }

\vspace{0.2cm}
{\it Bogoliubov Lab. Theor. Phys., JINR, Dubna, 141980, Russia}\\
{\it shirkovd@thsun1.jinr.ru}
\end{center}
\centerline{\bf Abstract }  
{\footnotesize
 The structure of the QFT expansion is studied in the framework of a new
``Invariant analytic" version of the perturbative QCD.  Here, an invariant
 (running)coupling $a(Q^2/\Lambda^2)=\beta_1\alpha_s(Q^2)/4\pi\,$, is
transformed into a ``$Q^2$--analytized" invariant coupling $a_{\rm an}(Q^2/
\Lambda^2)\equiv {\cal A}(x)\,$, which, by constuction, is free of ghost
singularities due to incorporating some nonperturbative structures. \par
 Meanwhile, the ``analytized" perturbation expansion for an observable $F$,
in contrast with the usual case, may contain specific functions ${\cal
A}_n(x)= \left[a^n(x) \right]_{\rm an}\,$, the ``n-th power of $a(x)$
analytized as a whole", instead of $\left( {\cal A}(x)\right)^n\,$. In other
words, the pertubation series for $F(x)$, due to analyticity imperative, may
change its form turning into an {\it asymptotic expansion \`a la Erd\'elyi
over a nonpower set} $\{{\cal A}_n(x)\}\,$. \par
  We analyse sets of functions $\{{\cal A}_n(x)\}\,$ and discuss properties
 of non-power expansion arising with their relations to feeble loop and
 scheme dependence of observables. \par
  The issue of ambiguity of the invariant analytization procedure and of
possible inconsistency of some of its versions with the RG structure
is also discussed.
}

\section{Introduction} %

In papers~\cite{jinr96,prl97} the procedure of {\sf Invariant Analytization}
of perturbative quantum chromodynamics (pQCD) has been elaborated. It
implements a {\sf combining} of two ideas: the {\sf RG summation} of leading
UV logs {\sf with analyticity property} in the $Q^2$ variable, imposed by
spectral representation\footnote{ This representation implements\cite{kniga}
 general properties of local QFT including the microscopic causality one. In
this context we shal use the term ``analyticity" as a synonym of
``spectrality" and ``causality".} of the K\"all\'en\cite{gunnar} --
Lehmann\cite{harry} type -- see Eq.(\ref{spectr}) below. This combination was
first proposed and devised~\cite{bls59} to get rid of the ghost pole in QED
about forty years ago.

  The central notion of the RG improved QFT perturbation technique is an
invariant charge or ``invariant coupling" $\,\alpha(Q^2/\Lambda^2)\,$
involved into the renorm--invariant expressions for observables. The QCD
scale parameter $\,\Lambda\,$ related to the ghost singularity position
equals approximately to 300 MeV. For properly normalized QCD invariant
(running) coupling $\,a(Q^2/\Lambda^2)=\beta_1 \alpha_s(Q^2)/4\pi\,$;
$\beta_1(n)=11-(2n_f)/3\,$ (defined in the space-like region), the
analytization results in a specific transformation into a form
$a_{an}(Q^2/\Lambda^2)$ free of ghost singularities. Here, by construction, \\
 {\it the analytic coupling is defined via the K\"all\'en--Lehmann
representation}
\begin{equation} a_{\rm an}(x)=
\frac{1}{\pi}\int\limits_{0}^{\infty}\frac{\rho(\sigma)\, d\sigma}{\sigma+x}
\; \; \; with \; the \; spectral \; density, \; \; \; \rho(\sigma)=  \Im
\,a(-\sigma)\, , \label{spectr}\end{equation}
{\it calculated on the basis of ``initial" RG-summed perturbation expression
$a(x)$ for invariant coupling.} \par \smallskip

   Generally, $a_{\rm an}(x)$ differs from its ``original input" $a(x)$
by nonpertur\-ba\-ti\-ve\footnote{ On a deep connection between
renormalization invariance plus causality and nonanalyticity in $\alpha$, see
our papers \cite{dv76-77}.} additive terms, which ``subtract" unphysical
singularities -- see, e.g., Eqs.~(\ref{a1an}) and (\ref{a2an}). \par

 A detailed analysis revealed \cite{jinr96,prl97} that the analytic coupling
$a_{\rm an}(x)$ obeys several important properties. It turns out to be
remarkably stable in the IR region, at $Q<\Lambda$, with respect to
higher-loop contribution and to renormalization scheme dependence. Its IR
limiting value $a_{\rm an}(0)$ is universal in this sense. We review this
subject quite shortly in Sections \ref{2-1} --- \ref{2-3}. \par

On the other hand, the Invariant Analytization of a physical amplitude
$F(Q,\alpha)$ is not a strightforward procedure. A few different scenarios
are possible. In papers \cite{MSS97,MSS97b}, a particular version, the
Analytic Perturbation Theory (APT), has been proposed and elaborated. Here,
due to specific analytization ansatz, instead of the power perturbation
series common for theoretical physics and QFT, an analytic amplitude
${\cal F}(x)$ is presented in a form of an asymptotic expansion of a more
general form, the expansion over an asymptotic set of functions
${\cal A}_n(x)=\left[a^n(x)\right]_{\rm an}\,$, the ``n-th power of $a(x)$
analytized as a whole". In the APT approach, the drastic reduction of loop
and renormalization scheme sensitivity for several observables has been found
-- see Refs.\cite{MSS97}--\cite{pl97}.

To understand the nature of the ``APT's loop and scheme immunity", in Section
\ref{3} we study properties of a nonpower asymptotic set
$\{{\cal A}_n(x)\}$ emerging from the APT recipe of analytization. \par

 In Section \ref{4}, we analyse the structure of possible variants of
analytization of expression for an observable and discuss the danger of
inconsistency (more precisely --- incompatibility with the inner structure
of RG) for some of them.

\section{RG solution and analytization\label{2}}

\subsection{One--Loop Analytization of $\,{\bf a(x)}$\label{2-1}}

At the one-loop level, the invariant coupling $\,a(x)=1/\ln x\,$ suffers from
a pole singularity at $\,x=1\,$ incompatible with the spectral representation
(\ref{spectr}). Here, analytization consists of analytic continuation of the
$\;1/\ln x\;$ expression into the negative $\,x\,$ region and defining the
spectral density via its imaginary part. The resulting spectral integral
(\ref{spectr}) with $\,\rho(\sigma)=\pi\cdot\left(\ln^2 \sigma
+\pi^2\right)^{-1}\,$ can be calculated explicitly
 \begin{equation} \label{a1an}
\label{tr1} a(x)= \frac{1}{\ln x} \, \Rightarrow \, a_{an}(x)={\cal A}(x)
= \frac{1}{\ln x} -\frac{1}{x-1} \, . \end{equation}

 The second term, precisely compensating the ghost pole, has a
nonperturbative nature. At small $\,a_{\mu}=a(Q^2=\mu^2)\,$ it is not
``visible" in Taylor series as it behaves like $\,\exp(-1/a_{\mu})\,$. To see
this clearly, one should return from $\,\Lambda$--parameterization to the one
in terms of $a_{\mu}$, the renormalized coupling constant, and $\,\mu^2\,$,
the reference momentum squared.

Note that a relation between $\,\Lambda/\mu$ and $a_\mu$ in the course of
analytization transformation changes. Instead of the usual expression
$\,\Lambda^2=\mu^2 \exp(1/a_\mu)$, according to Eq.(\ref{tr1}), we have the
 transcendental relation
$$\Lambda^2=\frac{\mu^2}{f(a)}\ ; \ \ \ a=\frac{1}{\ln f(a)}+
\frac{1}{1-f(a)}\, .$$
Here, at small $a$  (as well as in (\ref{a1an}) at large $x$), one can
neglect the second, nonperturba\-tive, term.  Meanwhile, at $\,a\simeq 1\,$
($x\leq 1)$ this term dominates, providing the IR fixed point at $a=1$.

The analytic coupling (\ref{a1an}) is a monotonous function in the whole
interval $(0,\,\infty)$ with the finite IR limit. The second term quickly
diminishes as $\,x\to \infty$: it contributes about $5 \%$ at $Q=10 \Lambda$
and only $1\%$ at $Q=25 \Lambda$. The whole set of solutions (\ref{a1an})
with various $\Lambda$ values, considered at the $Q^2$ scale, forms a bunch
with a common limiting point $a(0)=1$. This value, corresponding to
$\alpha_{\rm an}(0) = 4\pi/\beta_1$ \footnote{The effective flavour number at
residue of the pole, evidently is $n_f=3$. This  gives $\alpha_{\rm an}(0)=
1.396$.}, turns out to be universal. It does not change in the two- and
three-loop approximation as well -- see Refs. \cite{jinr96,prl97,ggk}.

\subsection{Two- and three-loop cases\label{2-2}}

 For the two-loop case, the invariant coupling has to be defined by
the transcendental relation
\begin{equation} \label{a-2}
\frac{1}{a^{(2)}(x)}-b\ln \left(1+\frac{1}{ba^{(2)}(x)}\right) =\ln x \ ;\;\;
b=\frac{\beta_2}{\beta^2_1}\,\ \left(=\frac{64}{81}\;\mbox{at}\ \ n_f=3\right)
\end{equation}
resulting from integration of the two-loop RG differential equation.

The iterative procedure yields the explicit approximate solution
\begin{equation} \label{a2iter}
\bar{a}^{(2)}_{\rm iter}(x)\,=\,\frac{1}{\ell+b\ln (1+\ell/b)}~ ,
\qquad \ell=\ln x \,  \end{equation}
used in our previous papers.

The exact solution to Eq.(\ref{a-2}) can also be expressed~\cite{badri,ggk}
\begin{equation} \label{lamb2}
a^{(2)}(x)=-\frac{1}{b}\cdot \frac{1}{1+{\cal W}(x)}\; ;\;\; {\cal W}(x)=
W_{-1}(z)\; ; \; \ z= - e^{-\ln x/b-1}\ \end{equation}
in terms of a special function $W$, the Lambert function
(we use the notation of paper \cite{LambertW})
$$ W(z) e^{W(z)}=z \, , $$
with an infinite number of branches $\,W_n(z)$.  Most part of the physical
region $\,x>1 \,$ corresponds to the particular branch $W_{-1}(z)$ that is
real and monotonous for $\,z<-1/e$. The relation between $z$ and $x$ in
Eq.(\ref{lamb2}) is normalized to correspond with approximation
(\ref{a2iter}) at $x\gg 1$. The branch ${\cal W}(x) = W_{-1}(z)$  is complex
below the ghost singularity at $x=1 \,, \,z=-1/e$.

The qualitative analysis of Eq.(\ref{a-2}) or its exact solution shows that
at $x=1+|\varepsilon|$ the ghost singularity has a form of the square-root
type branch point
$$
a^{(2)}(x\simeq 1)=\frac{1}{\sqrt{2b(x-1)}}-\frac{1}{3b}+O(\sqrt{x-1})\, $$
that yields an unphysical cut between this point and the origin.

To illustrate, proceeding from the two-loop solution (\ref{lamb2}), we define
the analytic coupling by the spectral representation (\ref{spectr}) with
spectral density defined in terms of ${\cal W}$, that is equivalent to
subtraction of the ``cut integral" related to the square-root singularity
\begin{equation} \label{a2an}
{\cal A}_2(x) =-\frac{1}{b(1+{\cal W}(x))}- \frac{1}{\pi}\int\limits_{0}^{1}
\frac{{\cal R}(\sigma)\, d\sigma}{\sigma-x}\, .
\end{equation}

 Note that iterative approximation (\ref{a2iter}) has a slightly different
structure of ghost singula\-ri\-ties\footnote{Besides the pole at $x=1$,
it has a logarithmic branch point at $x_c= e^{-b}$.}. \par
 Nethertheless, in Ref.\cite{badri} it has been demonstrated that the
analytized iterative solution is numerically very close\footnote{The $3 \div
4$ per cent difference can be ``compensated" by $-\, 8\%$ correction of the
 $\Lambda$ value.} to the analytized exact one, Eq.(\ref{a2an}). As a
practical result, this means that for the two-loop $a_{\rm an}(x)$ one can
use an expression in the form  Eq.(\ref{spectr}) with spectral density
 \begin{equation} \label{rho2}
 \rho^{(2)}_{RG} (L)\,=\,\frac{I(L)}{R^2(L)\,+\,I^2(L)}\, ;
\quad L=\ln\frac{\sigma}{\Lambda^2}\, ,  \end{equation}
$$
R(L)=L+b\ln\sqrt{\left(1+\frac{L}{b}\right)^2+\left(\frac{\pi}{b}\right)^2}
~, \; \; I(L)=\pi+b\,{\rm arccos}\frac{b+L}{\sqrt{\left(b+L\right)^2+\pi^2}}
~\, .$$

At the same time, in Ref.~\cite{ggk} it has been shown that the exact
three-loop solution (with Pad\'e transformed beta--function) can also be
expressed in terms of the Lambert function
\begin{equation} \label{lamb3}
a_3(x)=-\frac{1}{b}\cdot \frac{1}{1-c+W(z)}\, ; \ z=-e^{-\ln x/b+c-1}\  ;
\ c=\frac{\beta_3\beta_1}{\beta^2_2}\, .
\end{equation}
In what follows, referring to the three-loop case, we shall imply the
$\overline{\rm MS}$ scheme with $\beta_3^{\overline{\rm MS}}\,=\,2857/2
 -5033n_f/18 +325n^2_f/54\, .$
Here, at $n_f=3$, we have $ \beta_3= 3863/6=643.833 \ \ ; \ c = 1.415 \, .$

\subsection{Stability of analytic coupling in the IR region\label{2-3}}

The analytic coupling ${\cal A}(x)$ obeys several important properties in the
IR region. It is quite stable, at $Q<\Lambda$, with respect to higher-loop
contribution and to renormalization scheme dependence. Its IR limiting value
\begin{equation}\label{a=1}
\,{\cal A}(0)\,= \,1\, \end{equation}
(corresponding to $\,\alpha_{s,{\rm an}}(0)=4\pi/\beta_1(3)\simeq 1.4\,$)
is universal in this sense.    \par
 This remarkable property of universality was first noticed in our starting
up paper \cite{jinr96}. Later on, a detailed analysis
\cite{prl97,qcd97,badri,ggk} has revealed that the $\,{\cal A}(0)\,$ value
turns out to be insensitive not only to higher loop terms in the
beta-function but to the precise structure of the ghost singularity (removed
by analytization) as well. This structure depends on approximation.

  For instance, instead of the square-root singularity of the two-loop exact
solution (\ref{lamb2}), an iterative approximate solution Eq.(\ref{a2iter})
obeys a pole at $\,x=1\,$ and a log's branch point at $\,x_*=e^{-b}$. At the
three loop case, the Pad\'e--approximated solution obeys a pole and a branch
point (discussed in detail in the paper \cite{ggk}), to be compared with the
$(\ln x)^{-1/3}$ singularity of the solution with a non-transformed
beta-function.

 Nevertheless, in all these cases, the final analytic results for  $\,{\cal
A}(x)\,$ obey the property (\ref{a=1}) and their IR behavior in the interval
$\,(0,1)\,$ is very close~\cite{pl97} to each other. The analytization
procedure ``smoothing over all sharp angles" makes all them equal. \par
  One can also say that the spectrality bounds from above the invariant
coupling by this maximal value (\ref{a=1}), ``keeping it reasonably small".
For instance, usual beta-function in the ${\overline{\rm MS}}$ scheme for
$n_f=3$ numerically,
\begin{equation}  \label{beta3}
\beta(a)= -a^2(1+0.790 a+ 0.883 a^2) \sim -\alpha^2(1+ 0.566\alpha+
0.453\alpha^2)\, .
         \end{equation}
looks quite good: its higher terms are reasonably decreasing in the usual
physical region with $\alpha\leq 0.4\,(a\leq 0.3)$. It is worthwhile here to
introduce the ``beta-function for the analytic coupling"
\begin{equation} \label{beta-an}
\frac{d{\cal A}(x)}{d\ln x}=\beta_{\rm an}\left({\cal A}(x)\right)
  \end{equation}
which is a trancendental nonanalytic function $\,\beta_{\rm an}({\cal A})\,$
of its only argument (with the IR fixed point at $\,{\cal A}=1\,$). In the
one-loop case this function can be analyzed rather simply -- see, below,
Eqs.(\ref{A-2}) and (\ref{recurs1}). It turns out that its maximum value
$\,\beta_{\rm an}(1/2)= 1/12\,$ is of the same order of magnitude (this
statement remains valid in the higher loop case) as the expression
(\ref{beta3}) taken at $\,a=0.3\,$.

   Let us remind to the reader, that the connection between analyticity
of a function in the cut complex plane and its boundness have been discussed
several decades ago (see, e.g., Refs.\cite{tzu61}) in the context of a
low-energy behavior of hadron scattering amplitudes.
 \medskip

\subsection{Analytization of observables\label{2-4}}

  In papers \cite{MSS97,MSS97b} a specific recipe for analytization of
an observable $M(s)$ has been introduced. First, one should relate $M(s)$
possibly given in the time-like region, to some auxiliary function $f(Q^2)$
of a space-like argument\footnote{As, e.g., the $e^+e^-$ annihilation
cross-section ratio $R(s)$ is related to the Adler function $D(Q^2)$.}
which obeys the analyticity property in the $Q^2$ plane compatible with a
spectral representation of the K\"all\'en--Lehmann type. This function $f$,
after usual RG machinery, acquires a form of perturbative power series

\begin{equation} \label{8}
F\left(a(x)\right) =\sum_{n}^{} f_n a^n(x) \, .  \end{equation}

\parindent=4mm By prescription first used in paper \cite{MSS97}, the causal
analytization means
\begin{equation} \label{apt}
F\left(a(x)\right) \, \Rightarrow \, {\cal F}(x) =
\sum_{n}^{} f_n \,{\cal A}_n(x)\,
\end{equation}
with  ${\cal A}_n(x)=[a^n(x)]_{\rm an}\,$ being ``the n-th power of $a(x)$
analytized as a whole". Here, each  ${\cal A}_n(x)\,$ satisfies
the K\"all\'en--Lehmann representation with the spectral density
$\rho_n(\sigma)$ defined as $\Im a^n(-\sigma)$. \par

 Note that expansion (\ref{apt}) is not a power one as ${\cal A}_n(x)\neq
\left[{\cal A}(x)\right]^n$ at $n\neq1$. The recipe {\sf (\ref{8})$\,
\Rightarrow \,$ (\ref{apt})} changes the nature of expansion ! \par

  The surprise feature of this particular recipe called the ``Analytic
Perturbation Theory" (APT), is the remarkable stability of its results with
respect to a higher loop contribution \cite{MSS97,MSS97b} and, in turn, with
respect to a scheme dependence \cite{pl97}. We are going to show that the
origin of these physically important properties lies in the change of the
type of perturbation expansion.

The non-power set $\{{\cal A}_n(x)\}$ is an asymptotic one at UV as
$${\cal A}_{n+1}(x)\simeq [\ln x]^{-(n+1)}=o({\cal A}_n(x))\, , \ \ \, x
\to \infty \, . $$
 This means that the series (\ref{apt}), generally, should be treated as an
asymptotic expansion of the function $\,{\cal F}(x)\,$ over an asymptotic set
$\,\left\{{\cal A}_n(x)\right\}$. Its convergence features are determined, on
the one hand, by the coefficients $f_n$ calculated on the basis of $n$-loop
Feynman diagrams and, on the other, by the property of the set $\,\{{\cal
A}_n(x)\}$. Turn to the discussion of this set(s).

\section{Set of expansion functions\label{3}}

 \subsection{One-loop expansion functions\label{3-1}}

The simplest set $\{\,[a^n(x)]_{\rm an}\}=\{{\cal A}_n(x)\}$ consists of
analytized powers of the one-loop pQCD expansion parameter $a(x)$.
Besides (\ref{a1an}) it includes
\begin{equation}\label{A-2}
{\cal A}_2(x)=\frac{1}{\ln^2 x}-p_2(x)\, ; \ \ \,p_2(x)=
\frac{x}{(1-x)^2}=p_2\left(\frac{1}{x}\right) \, ; \end{equation}
$$
{\cal A}_3(x)=\frac{1}{\ln^3 x}+\frac{x}{(1-x)^3}-
\frac{1}{2}\frac{x}{(1-x)^2} \,;\;\; {\cal A}_4(x)=\frac{1}{\ln^4 x}-
p_2^2(x)-\frac{p_2(x)}{6}\,;\; \;\, \dots \;\; \,.$$

These ``analytized powers" obey a specific symmetry
$$
{\cal A}_n(x)=(-1)^n {\cal A}_n(1/x)\; ; \; (n>1)\,$$
and are related by recursion
relation with the help of the operator ${\cal D}= -x(d/d x)$
\begin{equation} \label{recurs1}
{\cal A}_{n+1}(x) = \frac{1}{n}{\cal D} {\cal A}_n(x)  =
\frac{1}{n!} {\cal D}^{n+1}\ln {\cal A}_0(x) \; \; \
\mbox{with} \ \ \ {\cal A}_0(x) = \frac{x-1}{x \ln x} \, \end{equation}
being the generating function.

In particular,
\begin{eqnarray}
\frac{d\,a_{an}(x)}{d\,\ln x} = - {\cal A}_2(x) \equiv \beta^{(1)}_{an}
\left(a_{an}(x)\right)= \beta^{(1)}_{an}\left(a_{an}(1/x)\right)\, .
\label{ beta-sym }\end{eqnarray}
Here, $\beta^{(1)}_{an}(a)$ is a non-analytic function of its argument.

An attempt to express ${\cal A}_n(x)$ via ${\cal A}(x)=a_{an}(x)$,
the analytic coupling, gives
\begin{equation} \label{mix}
{\cal A}_2(x)={\cal A}^2(x)-
\frac{2}{1-x}{\cal A}(x)+\frac{1}{1-x} \, ; \end{equation}
$$
{\cal A}_3(x)= {\cal A}^3(x) -\frac{3}{1-x}{\cal A}^2(x)+
\frac{3}{(1-x)^2}{\cal A}(x)-\frac{x+2}{2(1-x)^2} \, $$
---a sort of a ``mixed" representation, combining polynomial and
nonperturbative (via $x=Q^2/\Lambda^2$ argument)
dependencies. It can be argued (see below Section \ref{3-2}) that this
representation is not interesting from a pragmatic point of view.

In the two-loop case, to relate solution (\ref{a2an}) with higher analytized
powers, one can formally use the operator
\begin{equation} \label{d2}
{\cal D}_2=\frac{1}{b}\frac{\partial}{\partial {\cal W}}= -\frac{1+{\cal
W}(x)}{{\cal W}(x)}\cdot x\frac{\partial}{\partial x} \; ,\; \ \mbox{so that}
\; \; {\cal A}_{n+1}^{(2)}(x)=\frac{1}{n!}{\cal D}_2^n{\cal A}_n^{(2)}(x) \, .
\end{equation}

    Note also, that ${\cal D}$ in Eq.(\ref{recurs1}) can be treated as an
operator of differentiation over an ``effective time variable" $t=\ln x$.
An analogous interpretation of $\,{\cal D}_2\,$ gives
\begin{equation} \label{d2a}
 {\cal D}_2=\frac{1}{1+ba_2(e^t)}\cdot \frac{d}{d t} \equiv\frac{d}{d \tau}
\;  ; \;\;\; \tau = t+b\int_0^t a(e^{t'}) d t'  \,, \end{equation}
with $\tau(t)=t_2$, an ``effective two-loop time". For large $t$ values one
has $\tau\simeq t +b\ln t$. However, expression (\ref{d2}) needs to be
further specified at $\,0<x<1$.

\subsection{Subtraction Structures and behavior at ``the low ${\bf Q}$
           region" \label{3-2}}

The rational structures $p_n(x) ={\cal A}_n(x)-a^n(x)$ that subtract ghost
singularities
$$
p_1(x) =\frac{1}{1-x}\, ;\;p_2(x) =\frac{x}{(1-x)^2}\, ;\; p_3(x)
=\frac{x(1+x)}{2(1-x)^3} \, ;$$
 \begin{equation} \label{ss}
 p_4(x)=-\frac{x(x-x_+)(x-x_-)}{6(1-x)^4}\, ,\ \ (x_{\pm}=2\pm \sqrt{3})\ ;
\; \; \dots  \end{equation}
are  connected by a recursion relation analogous to (\ref{recurs1})
and, except $p_1$, obey (anti)symme\-t\-ry under $x \to 1/x \,$.

 As can be seen from this recursion relation, all $\,p_{n\geq 2}(x)$ at the
origin have the first-order zero that provides a property
\begin{equation} \label{1-0}
{\cal A}_n(0) =0 \ ;\ \ \ n\geq 2 \,
\end{equation}
valid in the higher loop case as well.

 For a quantitive orientation it is useful to study the ${\cal A}_n(x)$
behavior at $\,x=1+\epsilon:\;\epsilon \ll 1$. At the one-loop case this
can be done explicitly with the help of (\ref{ss})
$$
{\cal A}(x)\,\simeq\,\frac{1}{2}-\frac{\epsilon}{12}\,; \;
{\cal A}_2(x)\simeq\,\frac{1}{12} +\frac{19}{60}\epsilon^2\,;\; \,
\,{\cal A}_3(x)\,\simeq\, +\frac{\epsilon}{240} \, ; \;  \,
\,{\cal A}_4(1)\,\simeq\, -\frac{1}{720} +\frac{\epsilon^2}{3042}\, .$$

 These numerical results are rather instructive. Together with Eq.(\ref{1-0})
they show that in the ``low $Q$ region", at $Q\leq \Lambda$, we have
\begin{equation} \label{ll}
\left|{\cal A}_n(x)\right| \ll {\cal A}^n(x) \, , \end{equation}
an important estimate, which, at the very end, is responsible for a low
level of sensitivity of APT results with respect to the higher loop and
scheme effects -- see, e.g., observations made in Refs. \cite{MSS97} --
\cite{pl97}. The last estimate is valid also in the two- and three-loop cases.

 It is not practical to use expressions like (\ref{lamb2}), (\ref{lamb3}) and
(\ref{d2}) for explicit analysis of asymptotic sets $\{{\cal A}^n(x)\}$ at
two- and three-loop cases. For a short quantitative discussion we rather use
results of numerical calculation via an adequate spectral integral. They show
that in the interval $0\leq Q\leq\Lambda$ in addition to relation (\ref{ll})
we have
$$
{\cal A}_{k+1}(x)\cong \left[{\cal A}(x)\right]^{1+2k}\,.$$
 In other words, numerically, in the ``low $Q$ region" the real expansion
parameter is closer to ${\cal A}^2(x)$ rather than to ${\cal A}(x)$
\footnote{In particular, this means that a ``mixed" representation in powers
of ${\cal A}(x)$  with nonperturbative coefficients is not reasonable due
to big cancellation inside the r.h.s. of relations analogous to
Eq.(\ref{mix}).}. Naturally, as $Q/\Lambda$ grows and power terms diminish,
all ${\cal A}_n(x)$ tend to their natural limits $\left[{\cal
A}(x)\right]^n\,\simeq\,a^n(x)$. \par

 Moreover, as it follows from the representation (\ref{ss}) and property of
singularity $\ln^{-n}x$ at $x=0$, the expansion functions ${\cal A}_{n+2}(x)$
obey precisely $n$ zeroes on the interval $\left(0, X(\Lambda) \right)$ with
$X(\Lambda)$ being the upper boundary of the region where a power
nonperturbative correction $\sim x^{-1}= \Lambda^2/Q^2$ is essential. Hence,
the set under discussion consists of quasi-oscillating functions. This feature
makes the problem of estimating the resudial term (that is an error) in the
asymptotic expansion Eq.(\ref{apt}) more complicated. This is typical for
the {\it asymptotic expansion \`a la Erd\'elyi}.

Note also that the sets $\left\{{\cal A}_n(x)\right\}$ both in the one- and
two-loop cases obey a peculiar structure. Their neightbouring terms are
connected by differential relations (\ref{recurs1}) and (\ref{d2}).

\section{Discussion\label{4}}

    We have analysed a particular version of ``Causal, $Q^2$--analytic
perturbation theory", the APT version, which, by convention first introduced
in paper \cite{MSS97}, uses a set $\left\{{\cal A}_n(x)\right\}$ for
analytization of observables. It can be considered as a ``nonpower
analytization", to distinguish it from another possibility, the ``power" one
with the help of an asymptotic set $\left\{{\cal A}^n(x)\right\}$ by the
 recipe
\begin{equation} \label{power-an}
F\left(a(x)\right) \, \Rightarrow \, F\left({\cal A}(x)\right) =
\sum_{n}^{} f_n \left[{\cal A}(x)\right]^n\, ,
\end{equation}
instead of (\ref{apt}).  \par
 Just the nonpower analytization yields intriguing results with respect
to loop and scheme stability. At the same time, the power analytization,
Eq.(\ref{power-an}), results in a moderate change of usual pQCD practice
mainly in the IR region.

  This, technically simpler, second version has an advantage from a
theoretical point of view related to the issue of {\it Consistency of
analytization with the RG structure\,} -- see below Section \ref{4-2}. To
clarify, let us make a comment on the structure of the RG algorithm and on
``noncommutativity" of analytization with some of its elements. \par

\subsection{On ambiguity of analytization procedure\label{4-1}}

The procedure of the renorm-group method, in addition to deriving functional
and dif\-fe\-rential group equations, consists in a few steps\\
\noindent ${\bf [1]}$ {\sf Calculating beta--function(s) and anomalous
dimensions};\\
${\bf [2]}$ {\sf Solving RG differential equations (RGDEs) for invariant
coupling(s)} $a(x)$;\\
${\bf [3]}$ {\sf Solving RGDEs for other functions} $f(Q^2,\,a)$, e.g.,
propagator amplitudes, effective masses and ``physical amplitudes"\footnote{
Like , Adler functions, structure function moments, etc.}
{\sf with the use of explicit expressions for beta-function of Step {\bf [1]}
or invariant coupling(s) $a(x)$ obtained in Step {\bf [2]}}. The resulting
$F\left(a(x)\right)$ can be expressed as a power series (starting, possibly
with logarithmic term).

The invariant coupling analytization adds an additional step that follows
the Step ${\bf [2]}$: \\
${\bf [2a]}\hspace{45mm} a(Q^2)\to{\cal A}(Q^2)\,.\hspace{40mm}$

 However, analytization of propagators and observables can now be performed
either by modification of Step {\bf [3]} -- %
\smallskip

\noindent ${\bf [3m]}$ {\sf Using explicit expression ${\cal A}(x)$
in the process of RGDEs for $f(Q^2,\,a)$ solving } \par

or as an additional Step: \\
${\bf [4_{\rm APT}]}$ {\sf Analytizing the result of Step ${\bf [3]}$,
i.e., by applying the analytization procedure to the power series
for $F\left(a(x)\right)$} $ \Rightarrow {\cal F}(x)$.
\smallskip

The sequence
$${\bf [1]+[2]+[2a]+[3]+[4_{\bf\rm APT}] = [{\rm\bf APT}]} $$
was used in Refs.\cite{MSS97}--\cite{pl97}. Just this procedure
 yields nonpower asymptotic expansion (\ref{apt}).

On the other hand, in parallel with step ${\bf [4_{\rm APT}]}$ there exists
a simpler possibility: \\
${\bf [4_{\rm an}]}$ {\sf Substituting expression} ${\cal A}(Q^2)$,
like in (\ref{power-an}), {\sf in the result of Step ${\bf [3]}$.}\par

 The sequence
$${\bf [1]+[2]+[2a]+[3]+[4}_{\rm an}]= [{\rm\bf ICA}] $$ can also be used for
analytization of observables. This procedure, involving just the `invariant
coupling analytization' (ICA), yields power asymptotic expansion
(\ref{power-an}) differing from the usual one, Eq.(\ref{8}), by substitution
$a(Q^2/\Lambda^2)\, \Rightarrow \,{\cal A}(Q^2/\Lambda^2)$ only.

 We see that, generally, a causal analytization is not a unambiguous
procedure. Quite remarkably, the above--mentioned ambiguity is of a
functional nature (not in possibility to introduce an adjustable parameter).

\subsection{Analyticity vs RG structure ?\label{4-2}}

Meanwhile, the sequence ${\bf [1]+[2]+[2a]+[3m]}$ contains an inner
contradiction. E.g., Step ${\bf [3m]}$, used for the gluon propagator
amplitude\footnote{Compare with the Step ${\bf [4_{\rm an}]}$ used in
Ref.\cite{badri}.}, yields an expression that, at the very end, is not
compatible with the result of the previous Step ${\bf [2a]}$. At one loop
level, it gives \cite{sacha}
$$ d^i_{RG}(x) =\left[a(x)\right]^{\nu_i} \, \Rightarrow \,
\left[{\cal A}_0(x)\right]^{\nu_i} $$
with ${\cal A}_0(x)$ defined in (\ref{recurs1}). However, as it follows from
basic RG relations, the product of a vertex and appropriate powers of
propagators forms an invariant coupling. In the case under consideration,
one obtains  ${\cal A}_0(x)$  rather than $a_{\rm an}(x)={\cal A}(x)$ used
as an input. \par
   Quite analogously, there is a subtlety with the Step ${\bf [4_{\rm APT}]}$
implementation. The point is that for some objects (e.g., for propagator
amplitudes) the result of Step ${\bf [3]}$ at the one loop level starts with
fractional power (or logarithm) of $a(x)$ that gives rise to a branch point.
The analytization of expression like $\,[\ln x]^{-\nu}\,$ is equivalent to
subtraction of a cut contribution, i.e., yields a two-term structure of a
specific form\footnote{For explicit expressions we refer, e.g., to Refs.
\cite{bls59,badri}.}. It is easy to see that the appropriate product of such
structures cannot give expression (\ref{a1an}) for invariant coupling. Taken
literally, this observation means that the APT procedure also faces a
contradiction with the RG structure.
\bigskip %

{\Large\bf Conclusion}
 \smallskip %

\parindent=4mm 1. Our analysis in Section 3 reveals that the APT expansion,
Eq.(\ref{apt}), for an observable function, generally, represents an
asymptotic expansion over a nonpower asymptotic set $\{{\cal A}_n(x)\}$. The
latter obeys quite different properties in various ranges of the $x$
variable. In UV it is close to the power set $\{a^n(x)\}$, commonly used in
the current practice of QFT pertirbation calculation. Hence, the APT
converegence property in UV is completely determined by expansion
coefficients $f_n$.  On the other hand, in IR the asymptotic set $\{{\cal
A}_n(x)\}$ is of a more complicated structure.  In the ``low $Q$ region" the
behavior of the functions ${\cal A}_n(x)\,; n\geq 3 \,$ is oscillating. Due
to this, the contribution of higher terms in the APT expansion is suppressed.
The APT expansion, Eq.(\ref{apt}), in IR has a feature of asymptotic
expansion \`a la Erd\'elyi.  This tentative conclusion raises hopes that the
pertubative approach to QCD may be fruitful in the region $Q \sim 1$ GeV
where the QCD running coupling is not a small quantity.

 2. In Section \ref{4-1}, we have shown that the general program of Invariant
Analytisation, being quite a definite procedure for effective coupling,
is not ``rigid" enough when applied to other objects. In particular, it
contains a degree of freedom in analytizing observables.

   This ambiguity together with a ``proximity to contradiction", discussed in
Section \ref{4-2}, poses a question of looking for an additional ansatz in
the whole Causal analytic approach. The APT possibility is too interesting
to be ``abandoned without a struggle".

 3. In our opinion, one more funny lesson of the considered nonpower
construction is a semiquantitive observation that the APT approach is
equivalent to the usual pQCD practice with one strange amendment: ``To
restrict calculation to only the leading QCD contribution"; by the way,
forgetting about all headaches of higher-loop diagram calculation, scheme
dependency and expansion convergence. This intriguing feature could be
formulated as a suspicion that the pQCD is an effective theory\footnote{Like,
e.g., higher order perturbation contributions of the effective four-fermion
 Fermi weak interaction in QFT and of the BCS model Hamiltonian in the
theory of superconductivity.} and its higher order contributions have
no clear physical content.
\smallskip

{\large\bf Acknowledgements}
\vspace{0.1cm}

It is a pleasure to thank Dr. Igor' Solovtsov for important advices and help
in numerical calculations, as well as I.Ja. Arefeva, B.A. Arbuzov, V.S.
 Vladimirov, A.L.~Kataev, N.V. Krasnikov, B.A.~Magradze, A.V.Nesterenko and
A.A. Slavnov for useful discussion. Partial support by RFBR 96-01-01860,
 96-15-96030 and INTAS 96-0842 grants is gratefully acknowledged.


\begin{thebibliography}{50}
\bibitem{jinr96} D.V.~Shirkov and I.L.~Solovtsov, ``Analytic QCD running
	   coupling with finite IR behavior and universal $\alpha_s(0)$
           value", JINR Rapid Comm. No. 2[76]-96, 5-10, hep-ph/9604363.
\bibitem{prl97} D.V.~Shirkov and I.L.~Solovtsov,
     {\it Phys. Rev. Lett.} \, {\bf 79}, 1209-12 (1997), hep-ph/9704333.
\bibitem{gunnar} G. K\"allen, {\it Helv. Phys. Acta} {\bf 25} (1952) 417.
\bibitem{harry} H. Lehmann, {\it Nuovo Cim.} {\bf 11} (1954) 342
\bibitem{kniga}  N.N.Bogoliubov and D.V.~Shirkov, {\sf Introduction into the
	theory of quantized fields} Wiley--Interscience, N.Y., 1959, 1980 --
        see chapter ``Dispersion Relations".
\bibitem{bls59}  N.N.Bogoliubov, A.A.Logunov and D.V.~Shirkov,
            {\it Sov. Phys. JETP}\, {\bf 10} (1959) 574-581.
\bibitem{dv76-77} D.V. Shirkov, {\it Lett. Mat. Phys.} {\bf 1} (1976) 179-82;
             {\it Lett. Nuovo Cim.} {\bf 18} (1977), 452-456.
\bibitem{MSS97} K.A.~Milton, I.L.~Solovtsov and O.P. Solovtsova,
	       {\it Phys. Lett.} {\bf B 415},(1997) 104, hep-ph/9706409.
\bibitem{MSS97b} K.A.~Milton, I.L.~Solovtsov, and O.P. Solovtsova,
               {\it Phys.Lett} {\bf B 439} (1998) 421-427; hep-ph/9809510;
         ``The Gross--Llewellyn Smith sum rule in the analytic approach to
      perturbative QCD", Oklakhoma Univ. preprint OKHEP-98-07, hep-ph/9809513.
\bibitem{pl97} I.L.~Solovtsov and D.V.~Shirkov, {\it Phys. Lett.}\/
               {\bf B 442} (1998) 344-8, see also hep-ph/9711251.
\bibitem{badri} B.A.~Magradze, ``The gluon propagator in Analytic Perturbation
             theory", talk presented at the Intern. Conf. ``Quarks-98",
             Suzdal,  May 1998, hep-ph/9808247;
\bibitem{ggk} E.~Gardi, G.~Grunberg and M. Karliner, ``Can the QCD running
            coupling have a causal analyticity structure?", hep-ph/9806462;
             JHEP 07 (1998) 007.
\bibitem{LambertW} R.M.\,Corless et al., {\it Adv. in Comput. Math.} {\bf 5},
               (1996) 329.
\bibitem{qcd97} D.V.~Shirkov, {\it Nucl. Phys. B (Proc. Suppl.)}
             {\bf 64},  106-9 (1998), hep-ph/9708480.                    
\bibitem{tzu61} A.V. Efremov, D.V. Shirkov, and H.Y.\,Tzu,
              {\it Sov. Phys. JETP}\ {\bf 14} 432-7 (1962);
            {\it Scientia Sinica} {\bf 10} (1961) pp 812-36 -- see also
	    Section 11.3 in the monograph Ref.\cite{sms67}.
\bibitem{sms67}  D.V. Shirkov, V.A. Meshcheryakov and V.V. Serebryakov,
             {\sf Dispersion Theories of Strong Interactions at Low
             Energy}, North--Holland, 1969.
\bibitem{sacha} A.V. Nesterenko,The master thesis, Phys. Dept. of Moscow
                 State University, 1998.
\end{thebibliography}
\end{document}